\documentclass[aps,prb,twocolumn,superscriptaddress,amsmath,amssymb]{revtex4}
\usepackage{tabularx}
\usepackage{graphicx}
\usepackage{dcolumn}
\usepackage{bm}
\usepackage{color}

\newcommand{\bS}{{\vec S}}
\newcommand{\bR}{{\vec R}}
\newcommand{\bq}{{\vec q}}
\newcommand{\bQ}{{\vec Q}}
\newcommand{\bp}{{\vec p}}
\newcommand{\bk}{{\vec k}}
\newcommand{\bn}{{\vec n}}
\newcommand{\bL}{{\vec L}}

\newcommand{\pib}{\vec{\pi}}

\begin{document}

\title{Two-particle decay and quantum criticality in dimerized antiferromagnets}

\author{L. Fritz}
\affiliation{Institut f\"ur Theoretische Physik, Universit\"at zu K\"oln,
Z\"ulpicher Stra\ss e 77, 50937 K\"oln, Germany}
\author{R. L. Doretto}
\affiliation{ Instituto de F\'isica Te\'orica, Universidade Estadual Paulista,
01140-070 S\~ao Paulo, SP, Brazil}
\author{S. Wessel}
\affiliation{Institut f\"ur Theoretische Physik, Universit\"at Stuttgart,
Stuttgart, Germany}
\author{S. Wenzel}
\affiliation{Institute of Theoretical Physics, \'{E}cole Polytechnique F\'{e}d\'{e}rale de Lausanne (EPFL),
CH-1015 Lausanne, Switzerland}
\author{S. Burdin}
\affiliation{Condensed Matter Theory Group, LOMA, UMR 5798, Universit\'{e} de Bordeaux I,
33405 Talence, France}
\author{M. Vojta}
\affiliation{Institut f\"ur Theoretische Physik, Technische Universit\"at Dresden,
01062 Dresden, Germany}

\date{January 19, 2011}


\begin{abstract}
In certain Mott-insulating dimerized antiferromagnets, triplet excitations of the paramagnetic phase can
decay into the two-particle continuum. When such a magnet undergoes a quantum phase
transition into a magnetically ordered state, this coupling becomes part of the critical
theory provided that the lattice ordering wavevector is zero. One microscopic example is
the staggered-dimer antiferromagnet on the square lattice, for which deviations from O(3)
universality have been reported in numerical studies.
Using both symmetry arguments and microscopic calculations, we show that a non-trivial
cubic term arises in the relevant order-parameter quantum field theory, and assess its
consequences using a combination of analytical and numerical methods. We also present
finite-temperature quantum Monte Carlo data for the staggered-dimer antiferromagnet which
complement recently published results. The data can be consistently interpreted in terms
of critical exponents identical to that of the standard O(3) universality class, but with
anomalously large corrections to scaling. We argue that the two-particle decay of
critical triplons, although irrelevant in two spatial dimensions, is responsible for the
leading corrections to scaling due to its small scaling dimension.
\end{abstract}

\maketitle


\section{Introduction}

Coupled-dimer magnets have become model systems for quantum phase transitions
(QPT).\cite{ss_natph,ruegg_natph} Such dimerized magnets are obtained from placing
quantum spins on a regular lattice in $d$ spatial dimensions, with two spins per unit
cell and strong (weak) exchange interactions within (between) the unit cells. Depending
on the ratio of the exchange interactions, the ground state can be either a paramagnet,
dominated by singlet pairs in each unit cell, or a state with magnetic long-range order.
Experimentally, the QPT between these two phases can often be driven by pressure.
Additionally, these systems show another QPT upon applying an external field to the
paramagnet , which allows to realize Bose-Einstein condensation of
magnons.\cite{ruegg_natph}

On the theoretical side, it is commonly assumed that the zero-field QPT is -- by virtue
of quantum-to-classical mapping -- in the same universality class as that of the
$(d+1)$-dimensional classical Heisenberg model, often referred to as O(3) universality
class (note that the dynamical exponent $z=1$ in that case).\cite{ssbook} Numerical
simulations of various microscopic two-dimensional (2d) coupled-dimer Heisenberg models
have indeed found critical exponents consistent with three-dimensional (3d) O(3)
universality, in agreement with this
prediction.\cite{troyer97,matsumoto01,sandvik06,troyer08,wenzel09}
Therefore it came as a surprise when results from accurate
quantum Monte-Carlo (QMC) simulations of a particular coupled-dimer Heisenberg magnet,
the so-called staggered-dimer model with spins 1/2, displayed distinct deviations from
standard O(3) critical behavior, indicating a different universality
class.\cite{wenzel_prl}
In contrast, the often studied columnar dimer model was found to follow O(3) universality,\cite{matsumoto01,wenzel09}
suggesting the existence of two classes of coupled-dimer magnets.\cite{wenzel_thesis}
Subsequent QMC simulations of the staggered-dimer model,\cite{jiang09,jiang10,jiang10a} focusing
exclusively on the correlation-length exponent, obtained data
consistent with those of Ref.~\onlinecite{wenzel_prl}. However, it was argued that the
data for the largest systems could be
fitted to standard O(3) scaling laws.

In this paper, we propose a resolution to the puzzle provided by the numerical data. We
show that there are indeed two different classes of coupled-dimer magnets, henceforth
called A and B.
While class A follows standard O(3) universality, the low-energy quantum field theory of
class B is characterized by an additional cubic term which describes two-particle decay
of critical fluctuations and has no classical analogue. While similar cubic terms have
appeared before in the literature in different
contexts,\cite{parola89,einar91,fradkinbook,takano,mura} their effect on the critical behavior
has not been discussed to our knowledge. We also note that the two-particle decay of
triplet excitations at {\em elevated} energies has received some attention both
experimentally\cite{stone,masuda} and theoretically,\cite{kolezhuk06,zhito06} but it was
commonly assumed that such processes are negligible at lowest energies.
Here, we derive and analyze the critical two-particle decay term of class B in some
detail. While its precise characterization in two space dimensions presents a challenge,
our results are consistent with this term being weakly irrelevant in the
renormalization-group (RG) sense.

This leads us to suggest the following scenario for class-B coupled-dimer magnets like
the staggered-dimer model: The asymptotic critical exponents are the ones of the O(3)
universality class, but anomalously large corrections to scaling arise from the
two-particle decay term. This scenario is consistent with the numerical data reported in
Refs.~\onlinecite{wenzel_prl,wenzel_thesis,jiang09,jiang10,jiang10a}. In addition, we
also present finite-temperature QMC results for the temperature scaling of the quantum
critical uniform susceptibility, which lend further support to this scenario.

\subsection{Overview of results}
\label{sec:overview}

The QPT between a non-symmetry-breaking paramagnetic and a collinear antiferromagnetic
phase in an insulating magnet with SU(2) symmetry is typically described by a quantum field
theory of the $\phi^4$ type,
\begin{eqnarray}
\mathcal{S} \!&=&\! \int\!d^dr d\tau \frac{1}{2}\left[
c^2 (\vec{\nabla} \varphi_\alpha)^2 + (\partial_\tau \varphi_\alpha)^2 +
m_0 \varphi_\alpha^2 \right] + \frac{u_0}{24}
(\varphi_\alpha^2)^2
\nonumber\\
\label{phi4}
\end{eqnarray}
in standard notation.
Here, $\varphi_\alpha(\vec x,\tau)$ is a 3-component vector order-parameter field describing magnetic
fluctuations near the ordering wavevector $\vec Q$, with $\alpha=x,y,z$. For simplicity, the action has been
written for real $\varphi_\alpha$ (appropriate for time-reversal invariant $\vec Q$) and
isotropic real space; the generalization to other cases is straightforward. The critical
behavior of the model \eqref{phi4} is known to be of standard $(d+1)$-dimensional O(3)
universality.

We show that the following spatially anisotropic cubic term\cite{parola89,einar91,fradkinbook,takano,mura}
\begin{eqnarray}
\mathcal{S}_3 =
i \gamma_0 \int d^dr d\tau
\vec{\varphi}\cdot \left( \partial_x \vec{\varphi}\times \partial_\tau \vec{\varphi}\right),
\label{cubic}
\end{eqnarray}
with $x$ being a particular space direction, appears in the low-energy field theory for
2d coupled-dimer magnets belonging to class B. This term causes two-particle decay of critical fluctuations;
it bears some superficial similarity with Berry-phase and winding-number terms, to be
discussed below, however, its prefactor $\gamma_0$ is {\em not} quantized and the field
$\varphi$ is not restricted to unit length.

Our detailed analysis suggests that the cubic term $\mathcal{S}_3$ in $d=2$ space dimensions is
irrelevant in the RG sense, albeit with a small scaling dimension. It constitutes the
leading irrelevant operator at the critical fixed point. Consequently, the asymptotic
critical behavior is of O(3) type, but with anomalous corrections to scaling.
We show that this scenario is consistent with the existing numerical data.

\subsection{Outline}

The body of the paper is organized as follows:
In Sec.~\ref{sec:LMDM} we introduce the microscopic models under consideration, together
with the bond-operator representation of their Hamiltonian. We discuss the conditions for
the occurrence of cubic terms in the microscopic bond-operator formulation. Together
with the knowledge of the magnetic ordering wavevector, this allows to sub-divide the
models into classes A and B, where a cubic term does (B) or does not (A) occur in the low-energy
field theory.
Sec.~\ref{sec:field} is devoted to a careful derivation of this low-energy field theory for the
magnetic ordering transition in the presence of cubic terms of a specific model belonging to class B.
The critical behavior of this field theory will in turn be discussed in
Sec.~\ref{sec:scal}. We employ both scaling arguments and direct classical Monte Carlo
simulations to assess the relevance of the two-triplon decay term, with the conclusion
that the most plausible scenario is its weak irrelevancy.
This conclusion is supported in Sec.~\ref{sec:QMC} by QMC results obtained for
various coupled-dimer models of classes A and B.
A summary and outlook will close the paper.
Various technical details are relegated to the appendices.


\section{Lattice models of dimerized magnets}
\label{sec:LMDM}

In this paper, we consider Heisenberg models with spins 1/2, $\bS_j$, placed on a regular lattice
with spatially modulated couplings. The general Hamiltonian is thus
\begin{equation}
\mathcal{H} = \sum_{\langle jj'\rangle}J_{jj'}\bS_j\cdot\bS_{j'}
\label{ham}
\end{equation}
where the sum is over all pairs of lattice sites $jj'$.
Specifically, all models have two sites per unit cell, with an
antiferromagnetic intra-cell coupling $J'$ defining the dimers. Spins in different unit
cells are connected by couplings $J$ according to the underlying lattice geometry. We
restrict our attention to lattices without geometric frustration, where the classical
ground state is unique up to global spin rotations.
Various 2d examples, namely the staggered and columnar dimer models as well as the
herringbone and bilayer model, are shown in Fig.~\ref{fig:lattice}.

Quite generically, these coupled-dimer models possess a paramagnetic ground state for
$J'\gg J$, without symmetry breaking of any kind and dominated by intra-cell singlets
(Fig.~\ref{fig:lattice}). In contrast, for $J' \approx J$, a semiclassical N\'eel state
with broken SU(2) symmetry is realized. The critical properties of the resulting QPT as
function of $J/J'$ are the subject of this paper.

\subsection{Bond-operator representation}
\label{sec-bond}

An efficient microscopic description of the excitations of coupled-dimer models
is provided by the bond-operator representation.\cite{bondop}
Switching to a lattice of {\em dimer} sites $i$, the four states of a dimer $i$ can be represented using
bosonic bond operators $\{s_i^\dagger,t^\dagger_{i\alpha}\}$ ($\alpha=x,y,z$),
which create the dimer states out of a fictitious vacuum.
Explicitly (and omitting the site index $i$),
$|s\rangle = s^\dagger |0\rangle$,
$|\alpha\rangle = t^\dagger_{\alpha}|0\rangle$, where
$\left|s\rangle\right. =
(\left|\uparrow\downarrow\rangle\right.-\left|\downarrow\uparrow\rangle\right.)/\sqrt{2}$,
$\left|x\rangle\right. =
(-\left|\uparrow\uparrow\rangle\right.+\left|\downarrow\downarrow\rangle\right.)/\sqrt{2}$,
$\left|y\rangle\right. =
i(\left|\uparrow\uparrow\rangle\right.+\left|\downarrow\downarrow\rangle\right.)/\sqrt{2}$,
$\left|z\rangle\right. =
(\left|\uparrow\downarrow\rangle\right.+\left|\downarrow\uparrow\rangle\right.)/\sqrt{2}$.
The Hilbert space dimension is conserved by imposing the constraint
$s_i^\dagger s_i + \sum_\alpha  t^\dagger_{i\alpha} t_{i\alpha} = 1$ on every site $i$.
The original spin operators $\bS^1$ and $\bS^2$ of each dimer are given by
\begin{eqnarray}
S^{1,2}_\alpha &=& \pm\frac{1}{2}\left(s^\dagger t^{\phantom{\dagger}}_\alpha + t^\dagger_\alpha s
          \mp i\epsilon_{\alpha\beta\gamma}t^\dagger_\beta t^{\phantom{\dagger}}_\gamma \right),
\label{spin-bondop}
\end{eqnarray}
where $\epsilon_{\alpha\beta\gamma}$ is the antisymmetric
tensor with $\epsilon_{xyz} = 1$ and summation convention over
repeated indices is implied.

Using Eq.~\eqref{spin-bondop}, the Heisenberg Hamiltonian \eqref{ham} can now be re-written
in terms of the bond operators $\{s_i, t_{i\alpha}\}$.\cite{bondop,kotov,MatsumotoNormandPRL}
In the paramagnetic phase it is desirable to have a theory for triplet excitations only.
Among others, two routes have proven useful:
(a) In the spirit of spin-wave theory, the constraint is resolved by writing
$s_i=s_i^\dagger=(1-t^\dagger_{i\alpha} t_{i\alpha})^{1/2}$. Expanding the root then
generates a series of higher-order triplet terms.
(b) The formalism is re-interpreted as follows:\cite{kotov} Starting from a background
product state of singlets on all dimers, $|\psi_0\rangle = \prod_i s_i^\dagger
|0\rangle$, the operators $t^\dagger_{i\alpha}$ can be viewed as creating local triplet
excitations in the singlet background. This re-interpretation e.g. changes the local
triplet energy from $J'/4$ to $J'=J'/4 - (-3J'/4)$. The constraint takes the form of
a hard-core condition, $\sum_\alpha  t^\dagger_{i\alpha} t_{i\alpha} \leq 1$.
In both cases, subsequent approximations are usually designed to describe dilute triplet
excitations on top of the paramagnetic ground state.\cite{sommer}

\begin{figure}[t]
\includegraphics[width=8.5cm]{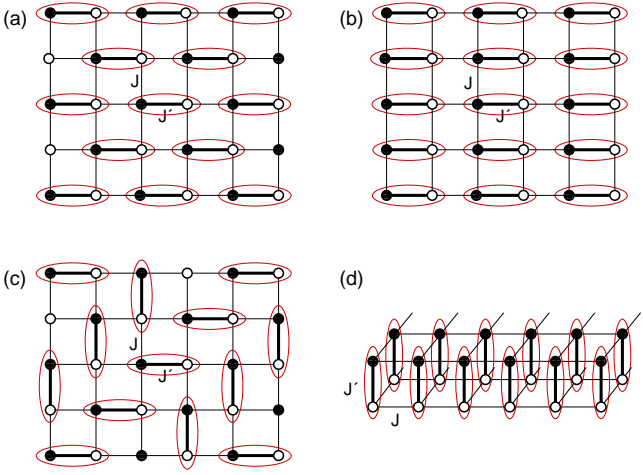}
\caption{ (Color online)
Two-dimensional coupled-dimer magnets considered in this paper. In all panels, thick (thin) bonds refer
to Heisenberg couplings $J'$ ($J$), solid (open) circles represent spins $\bS^1_i$
($\bS^2_i$) of each dimer. In addition, the singlet configurations in the paramagnetic
ground states realized for $J'\gg J$ are shown.
a) Staggered-dimer, b) columnar-dimer, c) Herringbone-dimer,
and d) bilayer Heisenberg model on the square lattice.
The QPTs to the antiferromagnetic phases are located at
a) $(J'/J)_c = 2.5196(2)$,\cite{wenzel_prl,wenzel09,jiang10,jiang10a}
b) $(J'/J)_c = 1.9096(2)$,\cite{wenzel_prl,wenzel09,sandvik10b}
c) $(J'/J)_c = 2.4980(3)$,\cite{wenzel_thesis}
d) $(J'/J)_c = 2.5220(1)$.\cite{wang06}
From the analysis in this paper, we conclude that the QPT of the models a) and c) belong
to class B, while that of b) and d) belong to class A, for details see text.
}
\label{fig:lattice}
\end{figure}

The final bond-operator Hamiltonian can be written as
\begin{equation}
\mathcal{H} = E_0 + \mathcal{H}_2 + \mathcal{H}_3 + \mathcal{H}_4 + \mathcal{H}_{\rm rest},
\label{ham-bond}
\end{equation}
where $E_0 = 3J'N/8$, with $N$ being the number of sites of the original (spin) lattice,
and $\mathcal{H}_{2,3,4}$ are terms obtained from $\mathcal{H}$ containing two, three,
and four triplet operators, respectively. The approaches (a) and (b) only differ in the
treatment of those higher-order terms in $\mathcal{H}_{\rm rest}$ which originate from the
different treatment of the Hilbert-space constraint: while (a)
leads to an infinite series (starting at quartic order), in case (b) $\mathcal{H}_{\rm
rest}$ only consists of the infinite on-site (i.e. hard-core) repulsion of
triplets.\cite{kotov}

Performing a Fourier transformation for the triplet operators, $t^\dagger_{i\,\alpha} = N'^{-1/2}\sum_\bk
\exp(-i\bk\cdot\bR_i)t^\dagger_{\bk\alpha}$, here on the lattice of dimers with $N'=N/2$,
$\mathcal{H}_{2,3,4}$ can be generically written as
\begin{eqnarray}
\mathcal{H}_2 &=& \sum_\bk A_\bk t^\dagger_{\bk\alpha}t_{\bk\alpha}
      + \frac{1}{2}B_\bk \left[ t^\dagger_{\bk\alpha}t^\dagger_{\bk\alpha}
        + {\rm H.c.}\right],
\label{hharmonic} \\
&& \nonumber \\
\mathcal{H}_3 &=& \frac{1}{2\sqrt{N'}}\epsilon_{\alpha\beta\lambda}\sum_{\bp,\bk}\xi_{\bk-\bp}
      \; t^\dagger_{\bk-\bp\alpha}t^\dagger_{\bp\beta}t_{\bk\lambda} + {\rm H.c.},
\label{hcubic} \\
&& \nonumber \\
\mathcal{H}_4 &=& \frac{1}{2N'}\epsilon_{\alpha\beta\lambda}\epsilon_{\alpha\mu\nu}
      \sum_{\bq,\bp,\bk} \gamma_\bk \;
      t^\dagger_{\bp+\bk\beta}t^\dagger_{\bq-\bk\mu}t_{\bq\nu}t_{\bp\lambda},
\label{hquartic}
\end{eqnarray}
with the coefficients $A_\bk$, $B_\bk$, $\xi_\bk$, and $\gamma_\bk$ depending on the
lattice geometry. Their explicit form for the models in Fig.~\ref{fig:lattice}a,b will be
given below.

In the paramagnetic phase, an expansion around the singlet product state $|\psi_0\rangle$
is justified. The leading-order term, $\mathcal{H}_2$, describes non-interacting triplet
excitations (``triplons''), with energy
\begin{equation}
\omega_\bk = \sqrt{A^2_\bk - B^2_\bk},
\label{energy-trip}
\end{equation}
obtained from a Bogoliubov transformation
$t^\dagger_{\bk\alpha} = u_\bk b^\dagger_{\bk\alpha} - v_\bk b_{-\bk\alpha}$
with coefficients
$u^2_\bk , v^2_\bk =  1/2 + A_\bk/2\omega_\bk,-1/2 + A_\bk/2\omega_\bk$ and $u_\bk v_\bk = B_\bk/2\omega_\bk$.
In the paramagnetic phase, $\omega_\bk > 0$ for all $\bk$.
Setting $\mathcal{H}\approx\mathcal{H}_2$ is often referred to as harmonic
approximation.

$\mathcal{H}_{3,4,{\rm rest}}$ contain interactions among the triplons. While
$\mathcal{H}_4$ and $\mathcal{H}_{\rm rest}$ will contribute to the quartic
self-interaction in the low-energy field theory, the cubic term $\mathcal{H}_3$ requires
a more detailed discussion: as will be shown below, it may induce a cubic term of the
form \eqref{cubic} in the low-energy theory, which then allows two-particle (in addition to
three-particle) decay of critical fluctuations.

\subsection{Bond operators for the staggered and columnar Heisenberg models}

\begin{figure*}[t]
\includegraphics[width=6.5cm]{dispersao-stag.eps} \hskip0.7cm
\includegraphics[width=6.5cm]{dispersao-col.eps}  \hskip0.5cm
\includegraphics[width=2.5cm]{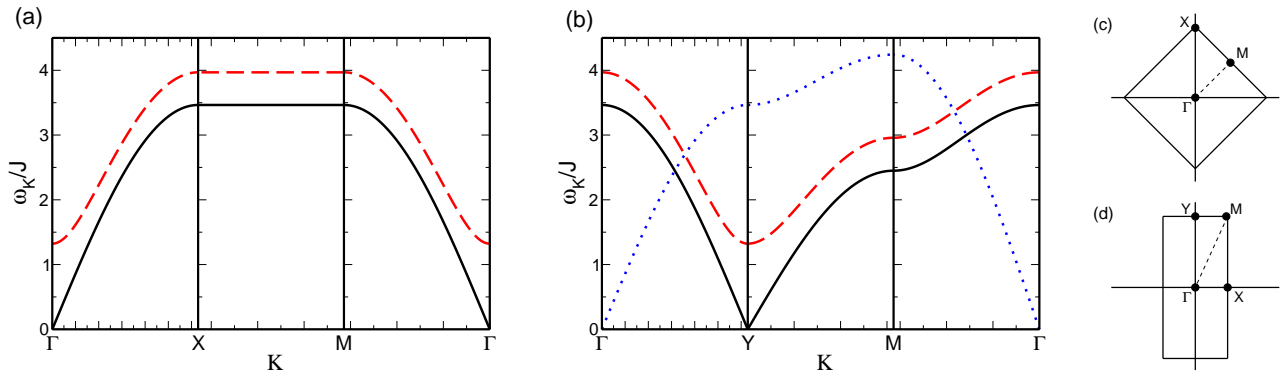}
\caption{(Color online) Triplet excitation spectrum (harmonic level) for the
  staggered (a) and columnar (b) dimerized AF Heisenberg models along
  high symmetric directions in the Brillouin zone. Solid (black)
  and dashed lines correspond respectively to $J'= 3J$ and $J'=
  3.5J$. The dotted line is the bottom of the two-particle
  continuum for $J'= 3J$ -- this coincides with the single-triplet dispersion in (a).
  Finally, panels (c,d) show the Brillouin zones for the staggered (c) and columnar
  (d) dimerized Heisenberg models.
}
\label{plot-energy}
\end{figure*}

For the staggered-dimer model in Fig.~\ref{fig:lattice}a, a straightforward calculation
gives
\begin{eqnarray}
 A_\bk      &=& J' + B_\bk,  \nonumber \\
 B_\bk      &=& -\frac{J}{2}\left[\cos(2k_x) + \cos(k_x+k_y) + \cos(k_x-k_y)\right], \nonumber \\
 \xi_\bk    &=& -J\left[\sin(2k_x) + \sin(k_x+k_y) + \sin(k_x-k_y)\right], \nonumber \\
 \gamma_\bk &=& B_\bk
\end{eqnarray}
where $k_{x,y}$ refer to momenta on the original square lattice of spins.
The triplon dispersion obtained at the harmonic level, i.e., from $\mathcal{H}_2$, is
shown in Fig.~\ref{plot-energy}a. Its minimum energy at $\bQ = (0,0)$ reaches zero at the
critical value $J'/J= 3$; the quantum Monte Carlo  result for the location of the QPT is $(J'/J)_c = 2.5196(2)$.
\cite{wenzel_prl,jiang10}

Similarly, for the columnar dimer model, Fig.~\ref{fig:lattice}b,
\begin{eqnarray}
A_\bk      &=& J' + B_\bk, \nonumber\\
B_\bk      &=& \frac{J}{2}\left[2\cos k_y - \cos(2k_x) \right], \nonumber \\
\xi_\bk    &=&  -J\sin(2k_x), \nonumber\\
\gamma_\bk &=& -\frac{J}{2}\left[2\cos k_y + \cos(2k_x) \right].
\end{eqnarray}
Now the dispersion minimum is at $\bQ = (0,\pm \pi)$, Fig.~\ref{plot-energy}b.
At the harmonic level, the critical point is again located at $J'/J= 3$, while
the currently most precise quantum Monte Carlo result is $(J'/J)_c = 1.90948(4)$,\cite{sandvik10b}
consistent with the previous value of $(J'/J)_c = 1.9096(2)$.\cite{wenzel09}

Fig.~\ref{plot-energy} also shows the lower bound of the two-particle continuum at the
critical coupling. In the columnar dimer model, Fig.~\ref{plot-energy}b, two-particle
decay is only possible at elevated energies (where the single-particle dispersion is
inside the two-particle continuum). In contrast, in the staggered-dimer model the
single-particle dispersion coincides with the bottom of the two-particle continuum at
criticality. While this coincidence at all wavevectors is an artifact of the harmonic
approximation, it is the correct result near $\bQ=0$ for a critical point with dynamical
exponent $z=1$. In the presence of a non-vanishing cubic term $\mathcal{H}_3$ in the
Hamiltonian, this implies a coupling between one-particle and two-particle sectors down
to lowest energies, as discussed in the following.

\subsection{Symmetries and two-particle decay}
\label{symmetries}

Before diving into the derivation of the low-energy field theory, it is worth discussing
for which coupled-dimer models a cubic term, describing two-particle decay of triplons,
will be part of the low-energy field theory.

First, a cubic piece $\mathcal{H}_3$ with $\xi_\bk\neq 0$ does not exist in all
microscopic models. In fact, in high-symmetry cases like the much-studied bilayer
Heisenberg model, Fig.~\ref{fig:lattice}d, $\xi_\bk= 0$.\cite{kotov}
An analysis shows that, due to the antisymmetric character of $\mathcal{H}_3$, its
coefficient $\xi_\bk$ is non-vanishing provided that two dimers $i,i'$ are coupled in an
asymmetric fashion, such that the couplings $J^{kk'}$ ($k,k'=1,2$) between the spins
$\vec S^k$ on $i$ and $\vec S^{k'}$ on $i'$ obey $J^{11} - J^{22}\neq 0$ and/or $J^{12} - J^{21}
\neq 0$. This can be translated into the following symmetry condition: $\xi_\bk$ vanishes
provided that the model remains invariant if in every dimer the spins 1 and 2 are
inter-changed (together with all their couplings).

Second, $\mathcal{H}_3$ can enter the low-energy theory only if the ordering wavevector
on the dimer lattice is $\bQ=0$ -- this immediately follows from momentum conservation.
As we shall show in the next section, these two conditions are indeed sufficient for a
non-vanishing cubic term \eqref{cubic} to appear in the low-energy theory, and thus these
conditions define our class B of coupled-dimer models.

Members of class B are the staggered-dimer model in Fig.~\ref{fig:lattice}a
as well as the coupled-dimer models on the honeycomb
lattice\cite{takano} and the herringbone square lattice,\cite{wenzel_thesis}
Fig.~\ref{fig:lattice}c -- these are exactly the models for which deviations
from O(3) critical behavior have been discussed.\cite{wenzel_prl,wenzel_thesis}
On the other hand, both the columnar and bilayer models, Figs.~\ref{fig:lattice}b and d,
do not fulfill these conditions and hence belong to class A. Indeed, standard O(3)
critical behavior has been established for these models.\cite{wang06,wenzel_prl,wenzel09}


\section{Low-energy quantum field theory}
\label{sec:field}

In this section we derive the effective low-energy field theory designed to capture
the physics near the antiferromagnetic quantum critical point of the models introduced in
Sec.~\ref{sec:LMDM}.
We shall present in detail the derivation of a $\varphi^4$ theory from the bond-operator
representation of the microscopic spin-1/2 model; in Appendix~\ref{app:nlsm} we shall
also sketch the derivation of a non-linear sigma model in the semiclassical limit of
\eqref{ham}. In both cases, a cubic term of the form \eqref{cubic} will appear for the
class-B models; however, we believe the non-linear sigma model is not useful for further
analysis, see Sec.~\ref{sec:cubic}.

\subsection{From bond operators to the $\varphi^4$ model}
\label{Sec:bondtophi}

A derivation of a $\varphi^4$ theory from the bond-operator formalism has been presented
in the context of the columnar dimer model in Ref.~\onlinecite{SachdevAF}. Here we shall
follow this procedure, taking also into account the cubic piece $\mathcal{H}_3$ which will
enter the $\varphi^4$ theory only if the ordering wavevector $\bQ=0$.

In a Lagrangian formulation, the bond operators can be represented by a complex bosonic
vector field $\vec t(\vec r,\tau)$. Importantly, this contains information both about the staggered
and uniform magnetization fluctuations on each dimer, i.e., it contains the degrees of
freedom of both $(\vec S^1-\vec S^2)$ and $(\vec S^1+\vec S^2)$. The latter live at high
energies and will eventually be integrated out to obtain a theory for the staggered
fluctuations only. To this end, we decompose the complex field
$\vec{t}=Z({\vec{\varphi}}+ia \vec{\pi})$, where ${\vec{\varphi}}$ and
${\vec{\pi}}$ are real three-component vectors, $a$ is the lattice spacing, and
\begin{eqnarray}
Z=2^{-\frac{3}{2}}a^{-\frac{3}{2}}J'^{\frac{1}{2}}
\end{eqnarray}
is a renormalization factor.

A continuum-limit formulation is obtained by an expansion of momenta in the vicinity of
$\bk = \bQ$. After some straightforward algebra the action corresponding to Hamiltonian
$\mathcal{H}_2$ \eqref{hharmonic} takes the form
\begin{eqnarray}\label{Eq:quadratic}
\mathcal{S}_2&=&\frac{1}{2}\int d^2 r d\tau
\left[c_x^2(\partial_x\vec{\varphi})^2 + c_y^2 (\partial_y\vec{\varphi})^2 + m_0 \vec{\varphi}^2
\right]
\nonumber \\
&+&\frac{1}{2}\int d^2r d\tau \left(
m_\pi \vec{\pi}^2+2i \sqrt{m_\pi} \vec{\pi}\partial_\tau \vec{\varphi}\right)
\end{eqnarray}
for both the staggered and columnar dimer models. Here, $m_0$ tunes the phase transition,
while $m_\pi$ remains finite near the QPT. Explicitly, we have $m_0
=J'\left(J'-3J\right)a^{-2}$ and $m_\pi=J'^2$. This implies that the transition at the
mean-field level takes place at $J'=3J$ (as in the harmonic bond-operator approximation
above). From the form of $m_0$ we can deduce its bare scaling dimension to be $2$
(relevant), since the continuum limit of the field theory obtains for $a \to 0$, and the
mass term grows upon approaching that limit. Furthermore, $c_x^2=3JJ'$ and $c_y^2=JJ'$.
The dynamics is encoded in the mixed-field linear time derivative.

The cubic Hamiltonian piece $\mathcal{H}_3$ \eqref{hcubic} of the staggered-dimer model,
Fig.~\ref{fig:lattice}a, can be cast into the form
\begin{eqnarray}\label{Eq:cubicpi}
\mathcal{S}_3 =\gamma_0 \int d^2r d\tau \, \partial_x \vec{\varphi}\cdot \left( \vec{\varphi}\times \vec{\pi}\right)\;,
\end{eqnarray}
where $\partial_x$ originates from $\xi_\bk \propto k_x$ for small $\vec k$, and
$\gamma_0 =J J'^{\frac{3}{2}}2^{\frac{1}{2}}a^{\frac{1}{2}}$. The latter implies that the
scaling dimension of $\gamma_0$ is $(-1/2)$, and consequently it is
irrelevant at tree level. Note that there is another term of the form
\begin{eqnarray}
\mathcal{S}'_3=\gamma'  \int d^2r d\tau \, \partial_x \vec{\pi}\cdot \left( \vec{\varphi}\times \vec{\pi}\right)\;.
\end{eqnarray}
This term turns out to be more irrelevant than $\mathcal{S}_3$ (it scales like
$a^{\frac{3}{2}}$) and hence will be discarded.

Importantly, a term of form $\mathcal{S}_3$ \eqref{Eq:cubicpi} does not appear for the
columnar dimer model, Fig.~\ref{fig:lattice}b: there, $\bQ = (0,\pm \pi)$, and a term
with three fields carrying momentum $\bQ$ is forbidden by momentum conservation. In
contrast, for other coupled-dimer models with ordering wavevector $\bQ=0$ and
non-vanishing $\xi_\bk$ in $\mathcal{H}_3$, a cubic term appears. We have checked this
explicitly for the honeycomb and herringbone lattices. The spatial structure of
$\mathcal{S}_3$ is strongly anisotropic, with the direction of the scalar spatial
derivative being determined by the orientation of the dimers (via the small-momentum
expansion of $\xi_\bk$). For instance, for the herringbone lattice,
Fig.~\ref{fig:lattice}c, this direction is diagonal w.r.t. the underlying square lattice.

The quartic interaction term is again identical for the staggered and columnar dimer
models and reads
\begin{eqnarray}
\mathcal{S}_4 =\frac{u_0}{24} \int d^2r d\tau\, \left[
( \vec{\varphi}^2 )^2+
\alpha_1 \left(\vec{\varphi}\times \vec{\pi} \right)^2 +
\alpha_2 \vec{\varphi}^2 \vec{\pi}^2\right],
\end{eqnarray}
where $u_0 =  3 \cdot2^{d+3} J'^3 a^{-1}$ and $\alpha_1,\alpha_2 \propto a^2$. Since the
terms $\propto \alpha_1,\alpha_2$ are more irrelevant than the first one we discard them.

The next step is to integrate out the field $\vec{\pi}$. Although $m_\pi>0$, care is required:
the coupling to $\vec\varphi$ via $\mathcal{S}_3$ renders the field $\vec\pi$ gapless at the critical point.
However, we have explicitly checked that this complication does not introduce singularities, see
Appendix~\ref{App:PI}.
To proceed, we introduce a new field
$\vec{\pi}'=\vec{\pi}+\frac{i}{\sqrt{m_\pi}}\partial_\tau \vec{\varphi}$, such that the
quadratic part of the action reads
\begin{eqnarray}
\tilde{\mathcal{S}}_2 &=&\frac{1}{2} \int d^2r d\tau
\left[c_x^2(\partial_x\vec{\varphi})^2 + c_y^2 (\partial_y\vec{\varphi})^2
      + (\partial_\tau\vec{\varphi})^2 + m_0 \vec{\varphi}^2 \right]
\nonumber \\
&+&\ \frac{m_\pi}{2} \int d^2r d\tau\, \vec{\pi}'^2
\end{eqnarray}
where now a second-order time derivative for $\vec{\varphi}$ appears.
Next, $\vec{\pi}'$ is integrated out from $\mathcal{S}_2+\mathcal{S}_3+\mathcal{S}_4$.
Keeping only the lowest-order terms and defining $\mathcal{S}_{24}=\mathcal{S}_2+\mathcal{S}_4$ we find
\begin{eqnarray}\label{Eq:LG}
\mathcal{S}_{24} \!&=&\! \frac{1}{2}\int d^2r d\tau
\left[c_x^2(\partial_x\vec{\varphi})^2 + c_y^2 (\partial_y\vec{\varphi})^2
      + (\partial_\tau\vec{\varphi})^2 + m_0 \vec{\varphi}^2 \right]
\nonumber \\
\!&+&\! \frac{u_0}{24}\int d^2r d\tau
\left ( {\vec{\varphi}}^2\right)^2
\end{eqnarray}
and
\begin{eqnarray}
\mathcal{S}_{3} &=& i \gamma_0 \int d^2r d\tau\,
\vec{\varphi}\cdot \left( \partial_x\vec{\varphi}\times \partial_\tau \vec{\varphi}\right)\;.
\label{Eq:cubicLG}
\end{eqnarray}
The additional term $\mathcal{S}_{3}$ \eqref{Eq:cubicLG} -- identical to \eqref{cubic}
announced in the introduction -- represents the crucial difference to a standard
$\varphi^4$ (or Ginzburg-Landau) theory as described by $\mathcal{S}_{24}$.
Among the various additional higher-order terms, there are
\begin{eqnarray}
\mathcal{S}_{3+2n}\propto i \int d^2r d\tau\, \vec{\varphi}\cdot \left( \partial_x \vec{\varphi}\times \partial_\tau \vec{\varphi}\right) \vec{\varphi}^{2n}
\end{eqnarray}
with $n>1$. As usual, they may be discarded since they are more irrelevant in the RG sense
compared to $\mathcal{S}_3$ \eqref{Eq:cubicLG}.

Summarizing, within the framework of the bond-operator approach one can derive, for the
staggered dimer model, an effective $\varphi^4$ theory supplemented by an infinite number of
additional terms. The most relevant of these is $\mathcal{S}_3$ \eqref{Eq:cubicLG}
which will be further discussed in Sec.~\ref{sec:cubic} below.

\subsection{Discussion of the cubic term}
\label{sec:cubic}

The derivation of the $\varphi^4$ theory has lead to a cubic term $\mathcal{S}_3$, Eq.
\eqref{Eq:cubicLG}, which is present for the staggered dimer model, but absent for the
columnar dimer model.

\subsubsection{Symmetries}

The cubic term \eqref{Eq:cubicLG} respects SU(2) symmetry and time reversal, as
$\vec\varphi$ is odd under time reversal. It respects momentum conservation provided that
$\vec\varphi$ parameterizes fluctuations at wavevector zero. Finally, the existence of
the term requires that $\vec\varphi$ is odd under the mirror operation $x\rightarrow -x$
while it is even under $y\rightarrow -y$, which is the case for the staggered
magnetization on the horizontally aligned dimers in the models in Fig.~\ref{fig:lattice}a,b.
The cubic term is fundamentally {\it quantum} in the sense that no quantum-to-classical
mapping exists for this term due to its prefactor $i$. (This also implies that the field
theory with cubic term is not amenable to an efficient Monte Carlo sampling, since it
suffers from a sign problem.)

\subsubsection{Quantization and relation to topological charge}

A cubic term can also be derived in the language of the non-linear sigma model, see
Appendix \ref{app:nlsm}, with the result
\begin{eqnarray}
\mathcal{S}_3 &=&  i \delta_0 \int d^d r d\tau \,
                   \bn\cdot(\partial_x\bn \times \partial_\tau\bn),
\label{comp-action}
\end{eqnarray}
where $\bn$ is now a unit-length O(3) field, and $\delta$ is proportional to the
modulation $(J-J')$ of the couplings in Fig.~\ref{fig:lattice}a. Similar cubic terms were
derived before for the standard\cite{parola89,einar91} and the dimerized\cite{takano}
Heisenberg models on the honeycomb lattice, but in all cases neglected in the subsequent
analysis.

Importantly, Eq.~\eqref{comp-action} may suggest an interpretation in terms of
a topological charge (or skyrmion number) in $x$-$\tau$ space. We can introduce a functional
$\mathcal{Q}$ of a vector field $\vec a$ in two dimensions according to
\begin{equation}
\mathcal{Q}[\vec a(x,y)] = \frac{1}{4\pi} \int dx dy \, \vec a\cdot(\partial_x\vec a\times \partial_y\vec a).
\end{equation}
For a unit-length field $\vec a$, $\mathcal{Q}[\vec a]$ is known as topological $\Theta$ term. It
is quantized to integer values for periodic boundary conditions and smooth configurations
of $\vec a$: $\vec a$ is a map from the 2d plane to the unit sphere, and $\mathcal{Q}$ measures how
often space is wrapped around the sphere.\cite{ssbook}
Notably, the Berry phase term $\mathcal{S}_B$ of a 1d antiferromagnetic spin chain,
represented in spin coherent states, can be represented in a very similar fashion:\cite{ssbook,tsvelik-book}
\begin{equation}
  \mathcal{S}_B = i\frac{S}{2}\int dx d\tau\,
                   \bn\cdot(\partial_x\bn\times \partial_\tau\bn)
                 = 2\pi i S\mathcal{Q}[\bn(x,\tau)].
\label{berry1d}
\end{equation}
Due to the quantization, $\mathcal{S}_B$ drops out from the partition function for
integer spins $S$ due to $e^{\mathcal{S}_B}=1$, while $\mathcal{S}_B$ contributes
non-trivial sign changes from skyrmions for half-integer $S$.\cite{haldane88} Now, the
dimer-model cubic term \eqref{comp-action} in $d=2$ can be written as:
\begin{equation}
\mathcal{S}_3 = 4\pi i \delta_0 \int dy \, \mathcal{Q}[\bn_y(x,\tau)]
\label{s3q}
\end{equation}
where $\bn_y(x,\tau) \equiv \bn(x,y,\tau)$. Note that, in contrast to $\mathcal{S}_B$ in
\eqref{berry1d}, the prefactor in $\mathcal{S}_3$ \eqref{s3q} is {\em not} quantized.
Based on the expression \eqref{s3q}, Ref.~\onlinecite{takano} concluded that
$\mathcal{S}_3$ is negligible, arguing that smooth configurations imply that
$\mathcal{Q}[\bn_y]={\rm const}$, and skyrmion lines described by non-zero
$\mathcal{Q}[\bn_y]$ should be energetically suppressed. This argument is certainly not
rigorous, as instanton events are not accounted for. However, the non-universal
prefactor of $\mathcal{S}_3$ may suggest that non-trivial contributions to
$\mathcal{Q}[\bn_y]$ tend to average out.

A central issue in the discussion of $\mathcal{S}_3$ is therefore whether the quantization in
terms of a topological charge in $x$-$\tau$ space really plays a role. We believe that
this is {\em not} the case, for the following reasons:
(i) In the $\varphi^4$ (i.e. soft-spin) version of the field theory, the field occurring
in $\mathcal{S}_3$ \eqref{Eq:cubicLG} is {\em not} normalized to unity, such that
$\mathcal{Q}[\vec \varphi_y]$ is not quantized. (Even if amplitude fluctuations are frozen
out at large length scales, $\mathcal{Q}[\vec \varphi_y]$ is sensitive to fluctuations on
all scales.)
(ii) Before taking the spatial continuum limit, the expression in $\mathcal{S}_3$
involves a discrete sum over $x$, with the derivative $\partial_x\vec\varphi$ replaced by a
linear function of $\vec\varphi$. While lattice definitions of $\mathcal{Q}$ preserving its
topological character for unit-length fields have been put forward,\cite{berg81} those
involve the fields in a strongly non-linear fashion. In contrast, for a discretization with a
linear approximation to the derivative it is easy to show that the topological character
is {\em not} preserved. This will be explicitly shown in Sec.~\ref{sec:mc} below.

Therefore, we believe that the cubic term in the order-parameter theory of class-B
coupled-dimer magnets is unrelated to quantized topological charges, i.e., the relation
suggested by Eqs.~\eqref{comp-action}, \eqref{s3q} is an artifact of the unit-length continuum limit underlying
the non-linear sigma model. Consequently, standard tools like perturbative RG can be used
to analyze the cubic term in the $\varphi^4$ formulation.


\section{Criticality in the presence of two-triplon decay}
\label{sec:scal}

According to our analysis so far, the cubic term $\mathcal{S}_3$ is the most relevant
additional term present in the low-energy field theory for the quantum phase transition
(as compared to the standard O(3) case). Therefore, the central question is whether this
term is relevant or irrelevant in the RG sense at the O(3) (or Wilson-Fisher) critical
fixed point in $(2+1)$ dimensions. This can be answered by determining the scaling
dimension of the coupling constant multiplying the local cubic operator which appears
inside $\mathcal{S}_3$.

In this section, we shall follow two routes:
First, we analyze the cubic operator in the $\varphi^4$ theory
perturbatively in $\epsilon=4-D$. Second, we determine the operator's scaling dimension
directly in $D=3$ dimensions, by means of a classical Monte Carlo calculation of the
operator's correlation functions at the Wilson-Fisher fixed point. The results of both
methods are consistent with the cubic term being weakly irrelevant in $D=3$.

\subsection{Perturbative determination of the scaling dimension}
\label{sec:scalphi4}

The O(3) critical fixed point is perturbatively accessible in the framework of the
$\varphi^4$ theory \eqref{phi4} in a double expansion in the quartic interaction $u_0$
and $\epsilon=4-D$. For $D<4$ the Gaussian fixed point is unstable towards the
Wilson-Fisher fixed point, with a renormalized interaction $u \sim
\mathcal{O}(\epsilon)$.

We start by determining the scaling dimension of the cubic operator's~\eqref{cubic} coupling constant at
the Gaussian fixed point. After re-scaling the lengths such that the gradient terms are
isotropic, the action in $D=d+1$ dimensions reads~\cite{wang06}
\begin{eqnarray}\label{Eq:LGnew}
\mathcal{S}&=&\frac{1}{2}\int d^Dr \left[
{m_0 \varphi_\alpha}^2 + (\vec{\nabla}\varphi_\alpha)^2\right]+
\frac{u_0}{4!} \int d^Dr\left(\varphi_\alpha^2\right)^2 \nonumber \\
&+& i\gamma_0 \int d^Dr \,
\vec{\varphi} \cdot \left(\partial_x \vec{\varphi}\times \partial_y\vec{\varphi}\right) \;.
\end{eqnarray}
At tree level, we obtain the well-known scaling dimensions
\begin{eqnarray}
~[\vec{\varphi}]_{\rm G}&=&(D-2)/2, \nonumber\\
~[u_0]_{\rm G}&=&D-4[\vec{\varphi}]_{\rm G}= 4-D, \nonumber\\
~[\gamma_0]_{\rm G} &=& D-2-3 [\vec{\varphi}]_{\rm G} = (2-D)/2,
\end{eqnarray}
with the subscript $G$ referring to the Gaussian fixed point. Substituting
$D=3$, the cubic term is found to be irrelevant with a scaling dimension of
$[\gamma_0]=-\frac{1}{2}$ -- the same conclusion appeared already in Sec.~\ref{Sec:bondtophi}.

At the Wilson-Fisher fixed point, both fields and vertices receive perturbative
corrections leading to anomalous dimensions. A simple (but incomplete) estimate of the scaling
dimension of $\gamma_0$ at the Wilson-Fisher fixed point consists of taking into
account the field renormalization only.
This amounts to using $[\gamma_0]= D-2-3[\vec{\varphi}]$ with $[\vec{\varphi}]=(D-2+\eta)/2$
leading to
\begin{eqnarray}
[\gamma_0] \approx \frac{2-D}{2}-\frac{3\eta}{2}\approx -0.55625
\label{scalgam}
\end{eqnarray}
where $\eta=0.0375(5)$ in $D=3$ (Ref.~\onlinecite{Campostrini}) was used.
Although indicative, we cannot expect this estimate to be reliable, as it ignores vertex
corrections: it is known that composite operators may have large anomalous
dimensions (see, e.g., Ref.~\onlinecite{SSdeconfined}).

A more complete treatment requires a perturbative RG analysis of the full theory
$\mathcal{S}_{24}+\mathcal{S}_{3}$. This expansion is done about the Gaussian
theory, with two dimensionless non-linear couplings $u=u_0\Lambda^{D-4}$ and
$\gamma=\gamma_0\Lambda^{\frac{D-2}{2}}$, where $\Lambda$ is an ultra-violet cutoff.
To one-loop order, the calculation is conveniently performed in the momentum-shell
scheme.
It turns out that, due to the antisymmetry of the $\gamma$ vertex, no diagrams mixing $u$
and $\gamma$ exist to one-loop order. Furthermore, $\gamma$ does not introduce field
renormalizations.
Hence, the flow equation for $u$ is not modified by $\gamma$, and the flow of $\gamma$
does not involve $u$. To one-loop order we simply have:
\begin{eqnarray}
   \frac{du}{dl} &=& (4-D)u - K_d \frac{N+8}{6} u^2, \\
   \frac{d \gamma}{dl} &=& \frac{2-D}{2}\gamma ,
\label{rgphi4}
\end{eqnarray}
where $dl=d\Lambda/\Lambda$, $N=3$ is the number of field components, and $K_d = (2^{d-1}
\pi^{d/2} \Gamma(d/2))^{-1}$. Thus, the tree-level result $[\gamma_0]=(2-D)/2$ does not
receive one-loop corrections.
If renormalizations of the $\gamma$ vertex due to $u$ remained absent at higher loop
orders, only field renormalizations would influence the flow of $\gamma$, and the
estimate \eqref{scalgam} would be correct. However, we see no fundamental reason for a
general cancellation of such vertex renormalizations.
Instead of going to higher loop orders, we will improve on the estimate
\eqref{scalgam} using a non-perturbative numerical approach.

\subsection{Monte Carlo analysis in $D=3$}
\label{sec:mc}

We shall now numerically determine the scaling dimension of the cubic term
$\mathcal{S}_3$ directly in $(2+1)$ dimensions at the Wilson-Fisher fixed point. Note
that this task is simpler than solving the full quantum model including $\mathcal{S}_3$:
in particular, it boils down to the simulation of a {\em classical} problem in $D=(d+z)$
dimensions, with $z=1$, as the O(3) critical field theory described by $\mathcal{S}_{24}$
follows a quantum-to-classical mapping.

We define the composite operator
\begin{eqnarray}\label{Eq:compositeoperator}
\mathcal{O}({\vec{r}})= \vec{\varphi}({\vec{r}})\cdot \left (
\partial_x \vec{\varphi}({\vec{r}})\times \partial_y \vec{\varphi}({\vec{r}}) \right)\;.
\end{eqnarray}
Its scaling dimension $\left[ \mathcal{O}\right]=\Delta_{\mathcal{O}}$ can be obtained
from the long-distance decay of its correlation function:
\begin{eqnarray}\label{Eq:correlator}
C({\vec{r}})=\langle \mathcal{O}({\vec{r}}) \mathcal{O}(0)\rangle \propto \frac{1}{|{\vec{r}}|^{2\Delta_{\mathcal{O}}}}.
\end{eqnarray}
From this, the scaling dimension of the coupling constant (more correctly: the associated vertex function)
is obtained through
\begin{eqnarray}\label{Eq:Gamma}
[\gamma_0] = D-\Delta_{\mathcal{O}} \;.
\end{eqnarray}

In the following, we determine the scaling dimension $\Delta_{\mathcal{O}}$ of the
composite operator $\mathcal{O}$ by a lattice Monte Carlo simulation of a classical
Heisenberg ferromagnet in $D=3$ dimensions, where we shall measure the correlator
Eq.~\eqref{Eq:correlator} at criticality. This approach exploits that the model is in the
same universality class as the O(3) Landau-Ginzburg theory and hence realizes the
Wilson-Fisher fixed point in $D=3$, but gives us access to correlation functions in a
non-perturbative manner. Specifically, we simulate the classical Heisenberg model
\begin{eqnarray}
H=-J\sum_{<ij>}\vec{S}_i\cdot \vec{S}_j
\label{cheis}
\end{eqnarray}
with ferromagnetic interactions between nearest neighbors on a simple cubic lattice.
The $\vec{S}_i$ are classical (commuting) three-component vectors of unit length
($\vec{S}^2_i=1$). We employ the Wolff cluster algorithm,\cite{wolff89} which allows an
efficient Monte-Carlo simulation and provides high-accuracy critical exponents for the
O(3) universality class.\cite{chen93} The critical point of this model is known to be
located at $K_c=J/(k_B T_c)=0.693035(37)$.\cite{chen93}

In the lattice simulation, the operator $\mathcal{O}$ needs to be discretized.
Guided by the derivation of the field theory from the discrete lattice model,
Sec.~\ref{sec:LMDM}, we know that the derivatives in Eq.~\eqref{Eq:compositeoperator}
should be discretized using a {\em linear} function of the spins (in contrast to
Ref.~\onlinecite{berg81}).
The standard two-point forward formula leads to
\begin{eqnarray}
\mathcal{O}_{i,\rm{lattice}}=\vec{S}_i\cdot \left(\vec{S}_{i+e_x}\times \vec{S}_{i+e_y} \right)\;,
\end{eqnarray}
where $e_x (e_y)$ denotes a unit step in x (y) direction. We have checked that other
(linear) discretization schemes give qualitatively similar results.

\begin{figure}[t]
\includegraphics[width=0.45\textwidth]{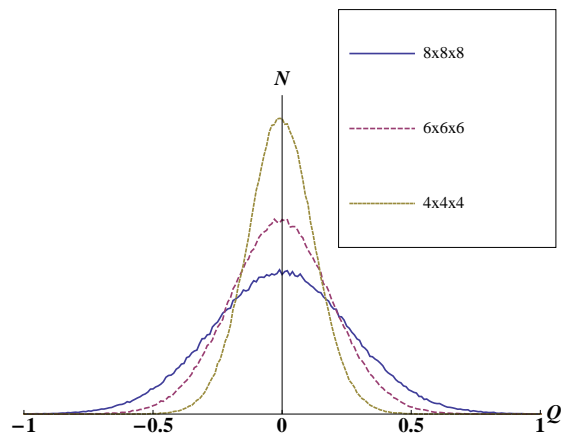}
\caption{Histograms of the intra-layer ``Skyrmion number'' $\mathcal{Q}(z)$ (see text),
obtained from a Monte-Carlo simulation of the classical Heisenberg model \eqref{cheis} at
its critical temperature.
The curves have been obtained from $10^8$ measurements on systems of size $L^3$ with $L=4,6,8$;
each measurement gave $\mathcal{Q}(z)$ for a single layer with fixed $z$.
Note that $\mathcal{Q}$ is not quantized, as it does not involve unit-length fields in the
continuum limit. The distributions are found to be Gaussian, with a width scaling linearly
with $L$.
}
\label{fig:skyrmion}
\end{figure}

Before we turn to the results for the correlator $C({\vec{r}})$, we make a brief detour
to discuss the quantity $\mathcal{O}$ itself. As mentioned above, its layer integral
$\mathcal{Q}(z) = \sum_{xy} \mathcal{O}(x,y,z) / (4\pi)$ may suggest an interpretation in
terms of a topological charge. However, the numerical Monte Carlo simulations show that
$\mathcal{Q}(z)$ is {\it not} quantized at criticality, see Fig.~\ref{fig:skyrmion}. Instead,
$\mathcal{Q}(z)$ displays a single peak at $\mathcal{Q}=0$, with a width scaling as $L$
in a system of size $L^3$. Since the number of spins in each layer is $L^2$, this width
simply reflects the standard thermodynamic scaling of fluctuations of a non-critical
extensive observable. Therefore, Fig.~\ref{fig:skyrmion} supports the conclusion of
Sec.~\ref{sec:cubic} that $\mathcal{O}({\vec{r}})$ is a conventional non-critical density.

\begin{figure}[t]
\includegraphics[width=0.48\textwidth]{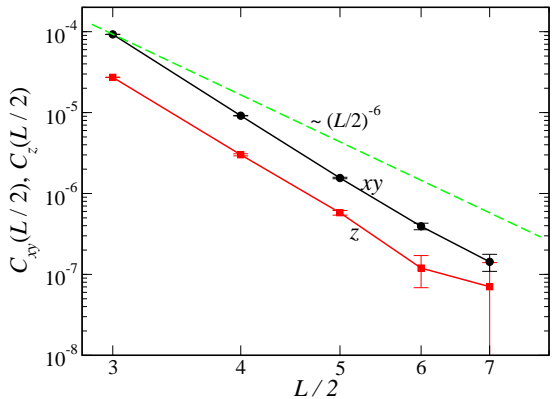}
\caption{
The correlators $C_{xy}(L/2), C_{z}(L/2)$ as functions of $L/2$ for different
system sizes $L$ from  Monte-Carlo simulations of the ferromagnetic Heisenberg model \eqref{cheis}
at its critical temperature. The dashed line corresponds to a decay
proportional to $(L/2)^{-6}$.
}
\label{fig:corr}
\end{figure}

The correlator of $\mathcal{O}$ is measured along the two inequivalent directions, i.e.,
within the $xy$-plane and along the $z$-axis:
\begin{eqnarray}
C_{xy}(r)&=&\langle \mathcal{O}(r,0,0) \mathcal{O}({\vec{0}})\rangle=\langle \mathcal{O}(0,r,0) \mathcal{O}({\vec{0}}) \rangle, \nonumber\\
C_{z}(r)&=&\langle \mathcal{O}(0,0,r) \mathcal{O}({\vec{0}}) \rangle
\end{eqnarray}
where $r$ now denotes discrete lattice coordinates.
We find that both correlation functions drop quickly with the separation $r$, and a large number of
Monte Carlo sweeps are required to reduce the statistical uncertainty in the correlator.
The most efficient way to estimate the decay exponents of $C(r)$ in a finite-size system is
to measure the correlation functions at half the linear system size, $C_{xy}(L/2)$ and
$C_{z}(L/2)$, for different lattice sizes, as to minimize finite size effects.
We employed up to $10^{12}$ Wolff cluster updates for system sizes $L=6,8,10,12,14$
to obtain the data shown in Fig.~\ref{fig:corr}.

Despite the relatively small system sizes, $C_{xy}(L/2)$ and  $C_{z}(L/2)$ show an
algebraic decay with $L/2$, consistent with critical behavior. For both correlators, the
decay appears to be faster than $1/r^6$.
However, a reliable determination of the decay exponent is difficult, because the data
display a slight curvature at the largest system sizes. A direct fit of all data points
yields $\Delta_\mathcal{O} \approx 4$, while the large-system data is more consistent
with $\Delta_\mathcal{O} \approx 3.2 \ldots 3.5$.
With Eq.~\eqref{Eq:Gamma} this suggests that the scaling dimension
$[\gamma_0]_{WF}$ is in the range $-0.2 \ldots -0.5$,
consistent with weak irrelevancy of the cubic operator
at the Wilson-Fisher fixed point.
However, from the present data we cannot rule out that $\gamma_0$ is instead
weakly relevant.

\subsection{Scenario: Large corrections to O(3) scaling}

Let us summarize the state of affairs concerning the critical behavior of class-B
dimer models:
(i) We have found that the combination of low symmetry and vanishing ordering
wavevector leads to the presence of a cubic term in the low-energy field theory -- this
cubic term represents the most relevant difference to a standard O(3) field theory.
(ii) The cubic term is strongly irrelevant in $(3-\epsilon)$ space dimensions, while the
Monte-Carlo results of Sec.~\ref{sec:mc} suggest that it has a small scaling dimension
in $d=2$, most likely being weakly irrelevant.
(iii) The published QMC results for the staggered dimer model indicate either critical
exponents slightly different from those of the O(3) universality class\cite{wenzel_prl}
or, if the fitting is restricted to large systems only, exponents consistent with their O(3) values.\cite{jiang09,jiang10}
Note that a conventional value for the critical exponent $\nu$ was obtained also using
an unconventional finite-size scaling analysis of the spin stiffness~\cite{jiang10a} whose validity
remains to be verified.

Points (i) and (ii) strongly suggest that the cubic term is responsible for the unusual
behavior seen in the QMC calculations. Point (iii) then implies that the cubic term is
{\em irrelevant} (instead of relevant) in the RG sense, as otherwise the deviations from O(3)
universality would grow (instead of shrink) with system size.

This leads us to propose the following scenario for the quantum phase transition in
class-B coupled-dimer models:
The cubic term is weakly irrelevant in $d=2$ and therefore constitutes the leading
irrelevant operator at the Wilson-Fisher fixed point. Hence, the asymptotic critical
behavior is that of the standard O(3) universality class, but the corrections to scaling
are {\em different} from standard O(3) universality. In the next section, we present
numerical results that support this scenario. A detailed analytical calculation of the
corrections to scaling arising from the cubic term is left for future work.

%
%
%


\section{Quantum Monte Carlo results}
\label{sec:QMC}

In this section, we present further results from QMC simulations of critical dimerized
antiferromagnets to assess the scenario discussed above. First, we re-analyze finite-size
data for critical exponents obtained in Ref.~\onlinecite{wenzel_prl} in terms of
anomalously large corrections to O(3) scaling. Second, we show results for the
finite-temperature uniform susceptibility at criticality -- this is a non-critical
quantity which has been studied both analytically and numerically for critical
antiferromagnets in the past.
Again, dimer models of classes A and B are found to display
distinctly different behavior.

\subsection{Correction exponent}

Following the idea of large corrections to scaling outlined above, we investigate here
whether the inclusion of appropriate scaling corrections yields critical exponents
compatible with the O(3) universality class for the staggered dimer model. In particular,
we focus on the order parameter (i.e. the staggered magnetization) $m_s$, estimated using
only the $z$-component of the spin operator via $m_s^z=|\sum_i (-1)^{x+y} S_i^z|$. The
analysis of this quantity lead to the most pronounced deviation from $O(3)$ critical
behavior in Ref.~\onlinecite{wenzel_prl}, and it complements more recent
simulations~\cite{jiang09,jiang10} that focus solely on the exponent $\nu$ of the
correlation length.
At the quantum critical point, $m_s^z$ is
expected to scale according to
\begin{equation}
  \label{eqn:dimer:plaq:beta}
  \langle m^z_s \rangle\sim L^{-\beta/\nu}(1+c_mL^{-\omega}),
\end{equation}
thus providing access to the ratio $\beta/\nu$ of critical
exponents. For the case of the ladder model, perfect agreement with
O(3) exponents was found --- even when neglecting the presence of
the corrections to scaling.\cite{wenzel09,wenzel_thesis}

For the standard O(3) universality class, the correction exponent $\omega$ is given
by\cite{hasenbuschO3} $\omega_{O(3)}=0.782(13)$ (in the $\varphi^4$ language arising from
the flow of the quartic interaction $u$ on the critical manifold). Hence, if the
deviations in $\beta/\nu$ observed in Ref.~\onlinecite{wenzel_prl} were due to standard
irrelevant operators, inclusion of $\omega=\omega_{O(3)}$ should result in the O(3) value
for $\beta/\nu$. Our results from performing fits to the data of
Ref.~\onlinecite{wenzel_prl}, presented in Table \ref{tab:dimer:staggeredbetacorr},
indicate that this is not the case, i.e., we cannot cast the fitting results with O(3)
values of the critical exponents.

We continue to investigate a second scenario, in which we fix the
known O(3) exponent $\beta/\nu$ but leave $\omega$ as a free fit
parameter.  The lower parts of Table~\ref{tab:dimer:staggeredbetacorr}
contain the corresponding fitting results, which indicate that we can
indeed arrive at a O(3) critical value for $\beta/\nu$, but at the
expense of a $\omega<\omega_{O(3)}$.  It should be understood that the
performed analysis actually provides an \emph{effective} correction
exponents for the length scales studied.
Nevertheless, from further numerical studies of the herringbone and honeycomb
coupled-dimer models,\cite{wenzel_thesis} we can empirically relate the presence of the
cubic operator to unusually large (and slowly vanishing) corrections to the leading O(3)
scaling. In fact, from the symmetry arguments presented in Sec.~\ref{symmetries}, both
models belong to class B (as the staggered-dimer model) with a cubic two-triplon decay
term at low energies. This provides numerical support for the scenario outlined
above, in which the cubic operator leads to enhanced corrections to an O(3) scaling
behavior.

\begin{table}
\centering
\begin{minipage}{\columnwidth}
  \caption{\label{tab:dimer:staggeredbetacorr}Fit results for
    the critical exponent quotient $\beta/\nu$ for the staggered dimer model. The table summarizes
    several results of fits including a correction to scaling exponent
    for three values of  $\alpha_\mathrm{c}$ within twice its error bar and
    $L\geq 10$. The entry (ref) refers to the relevant reference
    values for the 3D O(3) universality class given in the last
    line. Reported error bars are twice the fit error.}
  \begin{tabularx}{\textwidth}{XXXl}
    \hline\hline
    $\alpha_\mathrm{c}$ & $\beta/\nu$ & $\omega$ & $\chi^2/\mathrm{d.o.f}$ \\
    \hline
    $2.5194$  & $0.525(1)$  &  ref  &  $4.0$ \\
    $2.5196$  & $0.529(1)$  &  ref  &  $1.33$ \\
    $2.5198$  & $0.533(1)$  &  ref  &  $1.1$ \\
    \hline
    $2.5194$  & ref  & $0.63(3)$  &  $1.7$ \\
    $2.5196$  & ref  & $0.55(2)$  &  $0.6$ \\
    $2.5198$  & ref  & $0.48(3)$  &  $2.6$ \\
    \hline
    Ref.~\onlinecite{hasenbuschO3,guida} & $0.5188(3)$  & $0.782(13)$ & -- \\
    \hline\hline
  \end{tabularx}
\end{minipage}
\end{table}


\subsection{Finite-temperature susceptibility}

In this section, we focus on the thermodynamic behavior, in particular the temperature
scaling of the uniform susceptibility at the quantum critical point. For this purpose, we
performed QMC simulations of various dimerized two-dimensional antiferromagnets at
their respective quantum critical coupling ratios, see Fig.~\ref{fig:lattice}.

For the simulations, we employed the stochastic series expansion method with a directed
operator-loop update.\cite{sse1,sse2,sse3} We considered systems with $N=2L^2$ spin, for
linear system sizes $L$ up to 512. The  quantum non-linear sigma model prediction for the
uniform susceptibility is a linear dependence $\chi=A T$ on the temperature ($T$) within
the quantum critical region, where the prefactor $A$ depends on the spin-wave velocity
and a universal constant.\cite{chubukov93,chubukov94} This implies an essentially
constant ratio $\chi/T=A$ inside the quantum critical region. Such a linear-$T$ scaling
of $\chi$  has been observed for both the bilayer\cite{sandvik94} and a coupled plaquette
lattice\cite{troyer97} model significantly into the quantum critical region.

\begin{figure}[b]
\includegraphics[width=0.45\textwidth]{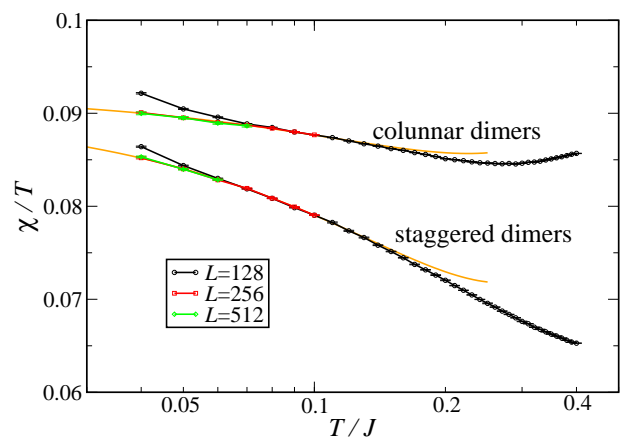}
\caption{
QMC results for the temperature dependence of the uniform susceptibility $\chi$ for the
columnar and the staggered  model at the quantum critical point for different system
sizes. Also shown are fits to an ansatz $\chi/T=A  - B T + C T^2$ for the low-$T$
behavior of $\chi/T$ extrapolated to the thermodynamic limit (dashed). Note the
logarithmic $T$ scale.
}
\label{fig:sus1}
\end{figure}

The QMC results of the uniform susceptibility for the staggered and the columnar dimer
arrangements are shown in Fig.~\ref{fig:sus1} for different system sizes. As seen from
comparing the QMC results for different system sizes, we obtain finite-size converged
estimates for the thermodynamic-limit behavior down to $T/J=0.04$. Comparing the data for
the two cases, we find that while for the columnar arrangement $\chi/T$ shows only mild
changes with $T$ below about $0.2 J$, for the staggered case, larger deviations from a
constant value of $\chi/T$ are observed over the whole accessible temperature range. We
take these enhanced deviations from the linear-$T$ scaling of $\chi$ as a signature of
the scenario outlined in the previous section, even though we are not in a position to
derive from our theoretical analysis the actual form of the leading deviation from the
linear-$T$ scaling of $\chi$.

\begin{figure}[t]
\includegraphics[width=0.45\textwidth]{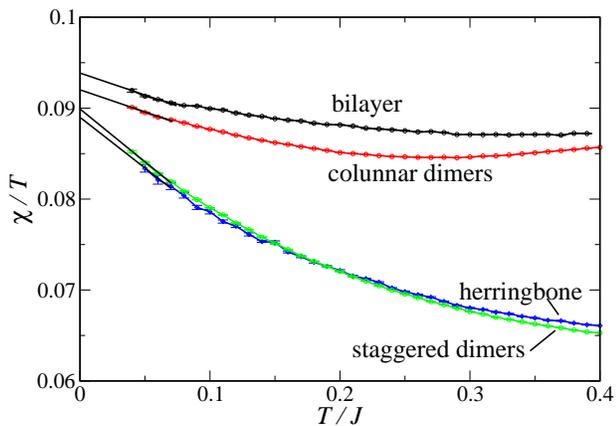}
\caption{
Quantum Monte Carlo results for the temperature dependence of the uniform susceptibility
$\chi$ for different dimer antiferromagnets. The shown data represents the thermodynamic
limit. Linear lines represent fits of the low-$T$ behavior of $\chi/T$ to a linear
ansatz.
}
\label{fig:sus2}
\end{figure}

In Fig.~\ref{fig:sus2}, we show the low-$T$ temperature dependence of $\chi/T$ for all
four models in Fig.~\ref{fig:lattice} at their respective quantum critical points,
extrapolated to the thermodynamic limit. Strikingly, we find sizeable and similar
corrections to the linear-$T$ scaling of $\chi$ for both class-B models (staggered dimer
and herringbone), whereas such corrections are much less pronounced for the class-A
models (columnar dimer and bilayer).

For all models, the leading low-$T$ behavior is consistent with a linear decrease of $\chi/T$,
i.e.,
\begin{equation}
\chi/T=A - B T,\quad B>0,
\label{eq:corr}
\end{equation}
shown by the linear fits in Fig.~\ref{fig:sus2}. For the staggered dimer and  the
herringbone model, enhanced corrections $B$ are required, as seen from the linear fit
lines in Fig.~\ref{fig:sus2}, which have about twice the slope as those for the other two
models. While polynomial corrections to the linear-$T$ scaling of $\chi$ are thus
compatible with our data, it is interesting to assess, if our data is consistent also
with other functional forms of the leading correction terms. For example, recently
Sandvik observed a dominating logarithmic term in  the low-$T$ corrections to the
linear-$T$ scaling of $\chi$ in a two-dimensional Heisenberg model with four-spin
interactions (the $J-Q$ model).\cite{sandvik10,sandvik10b}
For our data, the type of corrections can be judged from both Figs.~\ref{fig:sus1} and
\ref{fig:sus2}. Fig.~\ref{fig:sus1}, with a logarithmic $T$ scale, shows a quadratic fit
$\chi/T=A  - B T + C T^2$, which is seen to fit the data well up to about $T\approx
0.15$, while no robust behavior linear in $\log(T)$ is observed.

We thus conclude from our QMC analysis, that (i) at the quantum critical point deviations
from a linear-$T$ scaling of $\chi$ are exhibited by all considered models, (ii) the
leading corrections to the linear-$T$ scaling can be captured by a low-order expansion in
$T$ (Eq.~\ref{eq:corr}), and (iii) in those models, for which non-trivial cubic terms
emerge, considerably enhanced corrections to the linear-$T$ scaling of $\chi$ are
present. In fact, Figs.~\ref{fig:sus1} and \ref{fig:sus2} give a rather clear indication of the
two classes A and B of dimer models, with class A (B) displaying small (large)
corrections to the leading $\chi/T = {\rm const}$ behavior.
It proved difficult to extract the actual functional form of the scaling corrections from
the QMC simulations, but we tend to exclude low-temperature logarithmic corrections as
found for the $J-Q$ model in Refs.~\onlinecite{sandvik10,sandvik10b}.


\section{Conclusions}

Analyzing quantum phase transitions in models of coupled-dimer magnets, we have
identified two distinct universality classes A and B. While class A displays conventional
O(3) critical behavior, class B is characterized by the possibility of two-particle decay
of critical fluctuations, described by a cubic term in the order-parameter field theory.
We have shown that various 2d coupled-dimer models including the recently studied
staggered-dimer model belong to class B. Combining field-theoretic arguments and results
from large-scale numerical simulations, we have put forward the following scenario for
the quantum phase transition in class-B models: The leading critical behavior is that of
the standard O(3) universality class, but anomalously large corrections to scaling,
different from O(3) behavior, arise from the cubic term. This scenario appears consistent
with all available information, in particular it solves the puzzle concerning the
interpretation of recent QMC results for the staggered dimer
model.\cite{wenzel_prl,wenzel_thesis,jiang09,jiang10,jiang10a} A precise analytical
characterization of the scaling corrections arising from the cubic term is left for
future work.

It is conceivable that similar two-particle decay terms also appear for quantum phase
transitions with underlying symmetries different from SU(2). Then, the corresponding
class-B transitions might even display novel leading critical behavior.


\begin{acknowledgments}
We acknowledge useful discussions with W. Janke, A. Muramatsu, H. Rieger, A. Rosch, A. W.
Sandvik, D. Stauffer, E. Vicari, and especially S. Sachdev.
This research was supported by the DFG through the SFB 608 and SFB/TR 12 (K\"oln),
the SFB/TR 21 (Stuttgart) and the Research Unit FOR 960. Simulations have been performed
on the NIC J\"ulich and HLRS supercomputers.
\end{acknowledgments}


\appendix

\section{Derivation of $\varphi^4$ theory: self-energy corrections}
\label{App:PI}

In Sec.~\ref{Sec:bondtophi}, we derived a $\varphi^4$-type low-energy theory from the
bond-operator representation of the staggered dimer Heisenberg model. In the course of
the derivation we integrated out the field $\vec{\pi}$ based on the fact that it is
gapped. However, at the critical point this statement turns out to be true only at tree
level, due to the self-energy corrections to $\vec{\pi}$ arising from two-particle decay
described by Eq.~\eqref{Eq:cubicpi}.
Therefore we have to verify whether integrating out $\vec{\pi}$ is still permissible
without introducing singular terms.

Consider the self-energy of the field $\vec{\pi}$ due to the cubic
term~\eqref{Eq:cubicpi} which enables decay into two $\vec{\varphi}$ fields.
In lowest order perturbation theory and at criticality, it reads
\begin{eqnarray}\label{Eq:selfenergy}
\Sigma_{\vec{\pi}} \propto \frac{2}{3}\Lambda-\frac{\pi^2}{16}\frac{2
k_x^2+k_y^2+\omega_n^2}{\sqrt{k_x^2+k_y^2+\omega_n^2}} .
\end{eqnarray}
This implies that the tree-level gap is actually filled. Does this invalidate the step of
integrating out the field $\vec{\pi}$ as done in Sec.~\ref{Sec:bondtophi}? In order to
answer that question one has to estimate the emerging interaction terms. The most
relevant one of those is in the static limit given by
\begin{eqnarray}
\delta \mathcal{S} \propto \int d^dx d\tau \left(\vec{\varphi}^2\right)^2 \chi_{\vec{\pi}\vec{\pi}}({\vec{k}}=0,\omega_n=0)
\end{eqnarray}
with
\begin{eqnarray}
\chi_{\vec{\pi}\vec{\pi}}({\vec{k}}=0,\omega_n=0)\propto\int \frac{d^dq d\omega}{(2\pi)^{d+1}} G_{\pi}({\vec{q}},\omega)G_{\pi}(-{\vec{q}},-\omega)\nonumber \\
\end{eqnarray}
in which $G_{\pi}({\vec{q}},\omega)$ now is the full Green's function taking into account
the self-energy in Eq.~\eqref{Eq:selfenergy}. The respective integral reads
\begin{eqnarray}
\chi_{\vec{\pi}\vec{\pi}}({\vec{k}}=0,\omega_n=0)\propto\int \frac{d^Dq d\omega}{(2\pi)^{D}} \frac{1}{\left(m+\alpha \frac{2q_x^2+q_y^2+\omega^2}{\sqrt{q_x^2+q_y^2+\omega^2}} \right)^2}\nonumber \\
\end{eqnarray}
with $\alpha$ parameterizing the strength of the self-energy correction.
This expression reduces to case of gapped $\vec{\pi}$ for $\alpha=0$.
It is obvious that $\alpha \neq 0$ does not produce singular contributions.
We have performed similar checks for other terms generated from integrating out
$\vec\pi$.

We thus conclude that the presence of the cubic term ~\eqref{Eq:cubicpi}, although
rendering the field $\vec\pi$ gapless at criticality, does not invalidate the derivation
of the effective $\varphi^4$ theory \eqref{Eq:LG}.


\section{Derivation of the non-linear sigma model}
\label{app:nlsm}

Here we sketch the derivation of a non-linear sigma model for the staggered-dimer
Heisenberg antiferromagnet, as usual performed in the semiclassical limit starting from
the magnetically ordered phase (see Chap.~13 of Ref.~\onlinecite{ssbook}). A cubic term
will appear as a result of this derivation, further discussed in Sec.~\ref{sec:cubic}.

The Hamiltonian \eqref{ham} can be written as
\begin{equation}
\mathcal{H} = \sum_{j\delta}J_{j\delta}\bS_j\cdot\bS_{j+\delta},
\label{ham2}
\end{equation}
where $\delta=\hat{x},\hat{y}$ and
\begin{equation}
J_{j\delta} = J\left[ 1 + (\Delta/2)\delta_{\delta,\hat{x}}(1 + (-1)^j) \right] .
\label{j-stag}
\end{equation}
Here, $\Delta = J'/J - 1$ measures the modulation in the couplings,
and $(-1)^j = \pm 1$ for the two sublattices of the square lattice
(solid and open circles in Fig.~\ref{fig:lattice}a).

To derive a field theory, we replace $\bS_j \rightarrow S{\vec N}_j$, where ${\vec N}_j$
is a three-component unit-length vector. Assuming proximity to a state with collinear
N\'eel order, ${\vec N}_j$ can be parameterized by
\begin{equation}
 {\vec N}_j(\tau) = (-1)^j\bn_j(\tau)\left( 1 -
  \frac{a^{4}}{S^2}\bL^2_j(\tau) \right)^{1/2} + \frac{a^2}{S}\bL_j(\tau).
\label{parametrization}
\end{equation}
Here $\bn_i$ and $\bL_i$ are the (slowly varying) staggered and uniform components of the
magnetization, respectively, obeying the constraints $\bn_i^2 = 1$ and $\bn_i\cdot\bL_i =
0$. We have restored the lattice constant $a$ and assume that $\bL^2_i \ll S^2a^{-4}$.

Substituting Eq.~\eqref{parametrization} into the Hamiltonian \eqref{ham2}
and expanding the square root results in the following Hamiltonian piece $\mathcal{H}_y$
for the magnetic couplings along the $y$ axis:
\begin{eqnarray}
\mathcal{H}_y &=& JS^2\sum_j \left[ \frac{2a^{4}}{S^2}\bL^2_{j}
                              - \bn_j\cdot\bn_{j+\hat{y}}
          + (-1)^j\bn_j\cdot\bL_{j+\hat{y}}\frac{a^{4}}{S^2}\right.
\nonumber \\
  && \left.
         +\, (-1)^{j+\hat{y}}\bn_{j+\hat{y}}\cdot\bL_j\frac{a^{4}}{S^2}
         \right] + \mathcal{O}(L^3).
\label{ham-partial}
\end{eqnarray}
Upon taking the continuum limit, only the first two terms of Eq.~\eqref{ham-partial} are
finite. The other two terms oscillate on the lattice scale and disappear in the continuum
limit.
The remaining Hamiltonian piece, $\mathcal{H}_x$, is similar to $\mathcal{H}_y$ with
$\hat{y}\rightarrow\hat{x}$, with the crucial difference that the oscillating behavior of
the couplings changes the prefactors of the third and fourth term to $(-1)^j (-1)^j = 1$.
Therefore, these terms -- which will eventually lead to a cubic term analogous to
Eq.~\eqref{cubic} -- survive in the continuum limit for the staggered-dimer model (but
not for the columnar dimer model, simply reflecting that a cubic term is
forbidden by momentum conservation in the latter).

The continuum version of the Hamiltonian \eqref{ham2} then reads
\begin{eqnarray}
\mathcal{H} &=& \frac{J}{2}\int d^2rd\tau \,\Big\{ S^2\left[(\partial_y\bn)^2
       +  \left(1 + \Delta/2\right)(\partial_x\bn)^2\right]
\nonumber \\ \nonumber \\
  &&  + \, 4a^2(2 + \Delta/2)\bL^2 - 2Sa\Delta\bL\cdot(\partial_x\bn)
    \Big\}
\label{ham-cont}
\end{eqnarray}
for the staggered-dimer model, while the last term is absent in the columnar dimer model.

Passing to a coherent-state path-integral formulation and integrating out the $\bL$ fields,
the action for the unit-length $\bn$ field assumes the form
\begin{equation}
  \mathcal{S} = \mathcal{S}_B + \mathcal{S}_2 +  \mathcal{S}_3,
\label{action}
\end{equation}
where
\begin{equation}
\mathcal{S}_B = iS\sum_j(-1)^j\int_0^\beta d\tau \int_0^1 du\,
                \bn_j\cdot(\partial_u\bn_j \times \partial_\tau\bn_j)
\label{berry}
\end{equation}
is the familiar Berry phase term [Eq.~(13.52) of Ref.~\onlinecite{ssbook}]
and
\begin{eqnarray}
\mathcal{S}_2 &=& \frac{1}{2g}\int d^2r d\tau \,
                \left[c_x^2(\partial_x\bn)^2
                + c_y^2(\partial_y\bn)^2 + (\partial_\tau\bn)^2\right],
\nonumber \\
&& \nonumber \\
\mathcal{S}_3 &=&  i\frac{JS\Delta a}{g}\int d^2r d\tau \,
                   \bn\cdot(\partial_x\bn \times \partial_\tau\bn),
\label{comp-action2}
\end{eqnarray}
with
\begin{eqnarray}
   c_x &=& JSa\sqrt{8 + 6\Delta}, \nonumber \\
   c_y &=& JSa\sqrt{8 + 2\Delta}, \nonumber \\
     g &=& 2a^2J(4 + \Delta). \nonumber
\end{eqnarray}
The velocities $c_x$ and $c_y$ agree with the spin-wave velocities calculated within
spin-wave theory for the model \eqref{ham2}. The prefactor of $\mathcal{S}_3$
vanishes for $\Delta=0$, i.e., for the (unmodulated) square-lattice Heisenberg model.


\section{RG analysis of the non-linear sigma model}
\label{app:rgnlsm}

Despite the fact that the cubic term \eqref{comp-action2} in the non-linear sigma model
involves an artificial quantization as discussed in Sec.~\ref{sec:cubic}, one can
attempt a RG analysis of this non-linear sigma model using an expansion in $\epsilon =
(d-1)$. Here, we follow the calculation of Refs.~\onlinecite{nelson,chakravarty} and only display
the relevant changes.

We assume an isotropic velocity $c$ and a momentum-space ultraviolet cutoff $\Lambda$.
After re-scaling the coordinates according to $x_0 = \Lambda c \tau$ and ${\vec x}'=\Lambda{\vec x}$, the
action assumes the form
\begin{eqnarray}
    \mathcal{S} &=& \frac{1}{2g_0}\int_0^udx_0
           \int_1^\infty  d^dx \; \left[ (\partial_\mu\bn)^2  - 2hg_0n_z\right.
\nonumber \\
&& \nonumber \\
&&   \left. \;\;\;\;\;\;\;   -\, 2i\delta g_0 \bn\cdot(\partial_0\bn
                           \times \partial_1\bn) \right],
\label{action2}
\end{eqnarray}
where $\mu = 0,1,...,d$, $u = \Lambda c \beta$ is the re-scaled temperature,
$1/g_0 = \rho_S/(c\Lambda^{d-1})$ measures the stiffness $\rho_S$,
$h  =  H/(c \Lambda^{(d+1)}$ encodes an applied staggered field $H$, and $\delta =
\Delta/(c\Lambda^{d-1})$ represents the strength of the cubic term.
The tree-level scaling dimensions of the coupling constants follow as $[g_0] = 1-d$, $[h]
= d+1$,  and $[\delta] = d-1$. In contrast to the $\phi^4$-theory, here power-counting
indicates that the cubic term is relevant for $d \ge 1$.

The $\epsilon$ expansion is generated as usual via the parameterization $\bn =
(\pi_x,\pi_y,[1-\pi_x^2-\pi_y^2]^{1/2})$ and expansion of the action in $\pib$ fields.
Doing so, the lowest-order contribution of the original cubic term is found to be of
order $\mathcal{O}(\pib^4)$. Momentum-shell RG equations are obtained from integrating
out modes with momenta $e^{-l}<k<1$ and diagrammatically analyzing the perturbative
corrections as in Ref.~\onlinecite{chakravarty}. It turns out that the new coupling
$\delta$ does not modify the one-loop flow of $g$ and $h$, as the possible contributions
exactly cancel. The only correction to $\delta$ arises from a $\delta g$ diagram.

As a result, the one-loop RG equations in the limit $H \rightarrow 0$ and $T \rightarrow
0$ read:
\begin{eqnarray}
   \frac{dg}{dl} &=& (1-d)g + \frac{1}{2}K_d g^2,
\label{rg-eq1}\\
    \frac{d \delta}{dl} &=& (d-1)\delta + \frac{d+1}{d}K_d g \delta,
\label{rg-eq2}
\end{eqnarray}
where $K^{-1}_d = 2^{d-1}\pi^{d/2}\Gamma(d/2)$. Eq.~\eqref{rg-eq1} corresponds to the
limit $T \rightarrow 0$ of the equation (3.1a) from Ref.~\onlinecite{chakravarty}.

\begin{figure}[!b]
\centerline{\includegraphics[clip,width=4.0cm]{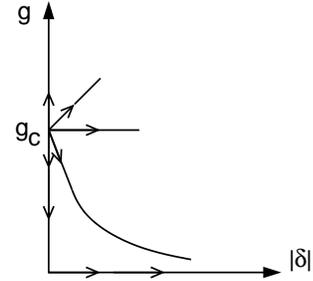}}
\caption{Schematic renormalization group flow for the nonlinear
  $\sigma$ model \eqref{action2} for the case $d>1$, $h=0$,
  and $T=0$.}
\label{rg-flow}
\end{figure}

The renormalization group flow is illustrated in Fig.~\ref{rg-flow}. The only non-trivial
fixed point is at $(g,\delta) = (g_c,0)$ with $g_c = 2(d-1)/K_d$. While this controls the
QPT for $\delta=0$, it is unstable w.r.t. finite $\delta$. While it is possible that the
inclusion of higher loop orders stabilizes a non-trivial fixed point at finite $\delta$,
the one-loop result in itself is puzzling:
Most disturbingly, the coupling $\delta$ becomes more relevant with increasing $d$
(already at tree level), in contrast to conventional expectations. One might argue that
in fact the combination $(g\delta)$, being marginal at tree level, measures the strength
of the cubic term. However, the absence of a stable fixed point remains to be understood.

Thus, the RG results for the non-linear sigma model (above) and for the $\varphi^4$ model
(Sec.~\ref{sec:scalphi4}) appear to mutually disagree regarding the role of the
two-particle decay term. As discussed in Sec.~\ref{sec:cubic}, we believe that the
non-linear sigma model analysis is not trustworthy, (at least partially) related
artifacts of the unit-length continuum limit. It is worth
mentioning that disagreement between the two field theories was already pointed out, for
instance, in Refs.~\onlinecite{pelcovits} and \onlinecite{castilla}. To our knowledge,
these issues are not completely settled.\cite{rmp-mirlin}



\end{document}